\documentclass[submission]{eptcs}
\usepackage{ulem}
\usepackage{graphicx}
\providecommand{\event}{TAV-WEB 2010} 
\usepackage{breakurl}             

\title{Preventing SQL Injection through Automatic Query Sanitization with ASSIST}
\author{Raymond Mui
\institute{Polytechnic Institute of NYU\\
6 Metrotech Center\\
Brooklyn, NY, 11201, USA}
\email{wmui01@students.poly.edu}
\and
Phyllis Frankl
\institute{Polytechnic Institute of NYU\\
6 Metrotech Center\\
Brooklyn, NY, 11201, USA}
\email{pfrankl@poly.edu}
}
\def\titlerunning{Preventing SQL Injection with ASSIST}
\def\authorrunning{R. Mui \& P. Frankl}
\begin{document}
\maketitle

\begin{abstract}
Web applications are becoming an essential part of our everyday lives. Many of our activities are dependent on the functionality and security of these applications. As the scale of these applications grows, injection vulnerabilities such as SQL injection are major security challenges for developers today. This paper presents the technique of automatic query sanitization to automatically remove SQL injection vulnerabilities in code. In our technique, a combination of static analysis and program transformation are used to automatically instrument web applications with sanitization code. We have implemented this technique in a tool named ASSIST (Automatic and Static SQL Injection Sanitization Tool) for protecting  Java-based web applications. Our experimental evaluation showed that our technique is effective against SQL injection vulnerabilities and has a low overhead.
\end{abstract}

\section{Introduction}

Web applications are essential parts of our everyday lives. They are becoming important and increasingly complex, as developers continuously add more and more features to enhance user experience. As web applications become more complex however, the number of programming errors and security holes in them increases, putting the users at increasing risk. Injection vulnerabilities such as SQL injection and cross-site scripting rank as top two of the most critical web application security flaws in the OWASP (Open Web Application Security Project) top ten list \cite{owasp}.

Web applications work as follows: a user requests a web page typically through a web browser. The request, which may include user inputs, is sent to the target web server through the HTTP protocol. 
The inputs become inputs to an application program
that is executed on the server side.
The program generates a new web page, which is sent back to the user, via HTTP.
Special inputs called cookies keep track of the current state between the user and the web server.
Many web applications interact with a database on the server side, in order to store persistent data that's relevant to the application,
such as user account information, product information, etc.
The web application program interacts with the database, building SQL statements and executing them through the database management system.

A problem with web application development is the threat of injection vulnerabilities such as SQL injection. This is caused by improperly sanitized user inputs. An example is shown in figure \ref{fig:code}, assuming the database contains a table \textit{BOOKS} with a string attribute \textit{author} and a numeric attribute \textit{price}. If the user selects \textit{author} as the action and enters \textit{\uline{John Doe}} then the SQL query \textit{SELECT * FROM BOOKS WHERE author = '\underline{John Doe}'} is constructed and sent to the DBMS and executed. However this code is vulnerable to SQL injection because the host variables from user inputs are not properly sanitized. If a malicious attacker enters the value \textit{\uline{';DROP TABLE BOOKS;$--$}}. The query becomes \textit{SELECT * FROM BOOKS WHERE author = '\underline{';DROP TABLE BOOKS;$--$}'} which causes the table to be dropped. This is an example of a SQL injection attack, as unsanitized user inputs caused a change in the structure of the intended SQL query.


\begin{figure} \scriptsize
    \begin{tabular}{ |p{15.5cm}| }
    \hline
      \begin{verbatim}
1.  String query = "SELECT * FROM BOOKS WHERE ";
2.  if(action == "author") {
3.       query += "author = '";
4.       query += getParam("author");
5.       query += "'";
6.  }
7.  else if(action == "price") {
8.       query += "price < ";
9.       query += getParam("price");
10. }
11. ResultSet rs = stmt.executeQuery(query);

\end{verbatim}\\ \hline

    \end{tabular}
  \caption{Motivating Example}
  \label{fig:code}
\end{figure}

Input injection vulnerabilities, such as vulnerabilities to SQL injection and cross-site scripting, can exist because of the way web applications construct executable statements, such as SQL, HTML, and Javascript statements, by mixing untrusted user inputs and trusted developer code. Currently the most widely used technique to prevent web application injection is requiring developers to perform proper input validation to remove these vulnerabilities. However, it is hard to do so because proper input validation is context sensitive. That is, the input validation routine required for the construction of SQL statements is different from the ones required for the construction of HTML, Javascript, etc. Because of this and the increasing complexity of web applications, manual applications of input validation are very error-prone. Just a single missed user input could lead to dire consequences.

The history of PHP's magic quotes \cite{magic_quotes} demonstrates the difficulty of proper input sanitization. It was intended as an automated measure against SQL injection by escaping all quotes from the user input. Simply escaping quotes however turns out to be a poor measure against SQL injection, and doing so causes issues when constructing statements of other languages like HTML and Javascript as well. In the end it caused more problems than it intended to solve and is being removed from the language altogether. Researchers have proposed several techniques for finding SQL injection vulnerabilities and for monitoring programs at run-time to prevent SQL injection. These are summarized in Section 6.

To help protect applications against SQL injection, we present a technique of automatic query sanitization. By using a combination of static analysis and program transformation, our technique automatically identifies the locations of SQL injection vulnerabilities in code and instruments these areas with calls to sanitization functions. This automated technique can be used to relieve developers from the error-prone process of manual inspection and sanitization of code. We have implemented our technique in a tool named ASSIST ({\bf A}utomatic and {\bf S}tatic {\bf S}QL {\bf I}njection {\bf S}anitization {\bf T}ool) for protecting
Java bytecode, which could come from applications developed as JSPs or Servlets.
Our experiments have shown that ASSIST is effective against a SQL injection attack test suite from Halfond et al. \cite{amnesia_testbed} and that our tool operates with a low runtime overhead. The main contributions of this paper are:

\begin{itemize}
\item {Our technique of automatic query sanitization of SQL queries through the use of static analysis and program transformation.}
\item {ASSIST: A proof of concept implementation of our technique for protecting Java Servlets and JSPs.}
\item {An experimental evaluation of ASSIST to demonstrate the effectiveness and performance of our technique.}
\end{itemize} 

The rest of this paper will be structured as follows: In section 2, we describe our technique and ASSIST in detail. In section 3, we provide an example walk-through of ASSIST, showing how it automatically instruments a vulnerable program step by step. In section 4, we discuss the possible limitations of this technique. In section 5, we show the results of our experimental evaluation, which demonstrates ASSIST's effectiveness at protecting against attacks and measures its performance overhead. In section 6, we discuss related work. In section 7, we conclude with a discussion of future work.

\section{Automatic Query Sanitization with ASSIST}

We now present our technique of automatic query sanitization. Our goal is to automatically insert calls to sanitization functions at appropriate points in the application code. There are two related issues:
\begin{itemize}
\item
Call location:
Where should the sanitization calls be placed?
\item
Call type:
Which sanitization function should be called at each such location?
\end{itemize}
Call location can be addressed by determining where variables that are data dependent on inputs are concatenated into strings that are eventually executed as queries. For call type, we need to examine the context in which those variables' values are used in the query. This is important because the choice of the sanitization function required for each variable depends on its corresponding attribute in the database schema. For example, in order to properly sanitize the code in figure 1, the sanitization function for the variable \textit{author} is different than the one for \textit{price}. This is because \textit{author} refers to a string attribute in the database in the query, while \textit{price} refers to an integer. This requires a more detailed analysis involving both the possible queries that can be executed at a given execution point and information about the database schema itself. 

The ASSIST algorithm has three phases.
\begin{enumerate}
\item
Find Query Fragments:
Static analysis is used to approximate the set of queries that can be executed at a given execution point.
\item
Parsing and Type Checking:
These sets of queries are then analyzed using a SQL parser and the database schema to determine the type of sanitization function need for each input.
\item
Instrumentation:
The code is transformed by adding calls to appropriate sanitization functions at appropriate locations.
\end{enumerate}

\begin{figure}
\includegraphics[width=6in]{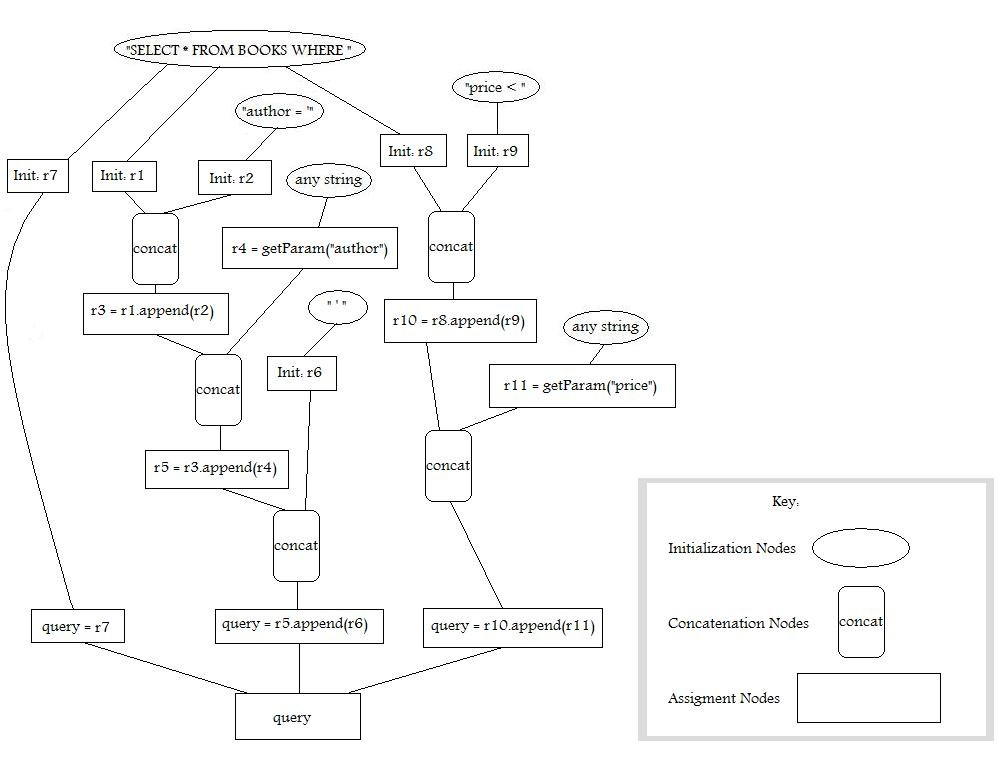}
\caption{Simplified string flow graph for the code in Fig~\ref{fig:code}}
\label{fig:flow_graph}
\end{figure}

Phases 1 and 3 use a {\it flow graph}, like the one used by Christensen et al~\cite{strings2003} in their Java String Analyzer\footnote{Java String Analyzer's output does not have all of the information needed for our analysis, so we work with its internal data structure, the flow graph.}.
The flow graph is a modified data flow graph where non-string statements are abstracted away. It keeps track of how string variables are created, assigned and modified throughout the entire program. The flow graph has three types of nodes: Initialization nodes, Assignment nodes, and Concatenation nodes. Initialization nodes represent initialization statements where a string local first gets created by the program. Initialization nodes contain the initial static values of literal strings when they are created or the value \textit{any\_string} for strings that get initialized from external sources, such as user input. Assignment nodes represent an assign statement where the value of one string variable gets assigned to another variable. Concatenation nodes represent a string concatenation between two string variables. Figure \ref{fig:flow_graph} shows a (simplified) flow graph for the example in figure \ref{fig:code}. Relevant string variables are shown in the assignment nodes. Variables $r1$, $r2$, etc. correspond to source variables or temporaries.

\subsection{Find Query Fragments}

We wish to determine the values
a string variable $query$ can have at a given query execution point and
the sources of input variables that are used in those queries.
We define an {\it abstract query} to be a sequence of string literals
and place-holders which represent a SQL query. The place holders represent nodes in the flow graph
corresponding to user inputs whose values are used in query construction.
A {\it query fragment} is a substring of an abstract query.
The Find Query Fragments
algorithm associates a query fragment set, $QFS(n)$
with each node $n$ in the flow graph. We define $QFS(n)$ recursively as shown in figure \ref{fig:qfs}. Computing QFS of the node representing a query execution point yields the possible abstract queries at that particular query execution point. 


\begin{figure} \scriptsize
\begin{tabular}{ | p{15.5cm} | }
    \hline
      \begin{verbatim}
Query-Fragment-Set FindQueryFragments(Node n)
{
  if n previously visited, return QFS(n) //previously computed
  mark n visited;
  case (type of n)
  {
    initialization node (string literal s):
      QFS(n) = s
    initialization node (any string):
      QFS(n) = place-holder reference to node n
    assignment node:
      QFS(n) = Union over p in pred(n) FindQueryFragments(p)
    concat node:
      //concatenate each element of left predecessor with each element of right predecessor
      QFS(n) = FindQueryFragments(left(n)) concat FindQueryFragments(right(n));
  }
  return QFS(n)
}
\end{verbatim}\\ \hline

    \end{tabular}
  \caption{Find Query Fragments Algorithm}
  \label{fig:qfs}
\end{figure}

First, our algorithm finds the nodes from the flow graph that represent the query execution point. Query execution points can be found at assignment nodes that contain calls to the \textit{java.sql} library functions that executes a SQL query, such as \textit{Statement.executeQuery(query)}. Starting from these nodes, we find their possible queries by calling FindQueryFragments on those nodes, which recursively computes the QFS of all their predecessors as follows:
The QFS of an initialization node with value ``any string" is a place-holder
pointing to that node.
The QFS an Initialization node representing a string literal is the string literal. The QFS of an Assignment node is the union of the query fragment sets from the nodes of its predecessors. The QFS of a Concatenation node is the QFS from the node on its left hand side concatenated with the QFS from the right hand side\footnote{
For efficiency, but with some potential loss of precision, our implementation does not concatenate every combination of left hand side and right hand side because that could lead to combinational explosion, however our implementation ensures that every value from each set from each side is concatenated in the result at least once.}. 
The algorithm marks nodes that have been visited to guarantee termination in the case that the flow graph has cycles.
As discussed in Section 4, this introduces some imprecision.
In addition,
the QFS of each node is memoized when the node is visited.
This eliminates some redundant computations.

\subsection{Parse Abstract Queries}

In order to determine the types of sanitization functions that are needed for variables at each place-holder marking node $n$, the abstract queries in QFS(execution point node) are analyzed with schema information. This is done by the parsing phase where the attributes they refer to from the query are matched with type information from the schema. We currently focus on the two types of sanitization functions -- string sanitizers and numeric sanitizers. More generally, one could define special sanitization or checking functions for each attribute domain in the database schema. For example, if an input variable is being compared to an attribute of domain CHAR(10), one could insert a sanitization function that checks that the input string does not have more than 10 characters.

In addition to determining the contexts in which user inputs are used in queries, the parsing step allows us to eliminate some syntactically incorrect queries from consideration. These may include some queries that are constructed along infeasible paths.

\subsection{Instrumention of Code}

The final step is instrumenting the code with calls to sanitization functions.
Given an abstract query $q$ with a place-holder marking node $n$ (i.e. showing that an input from node $n$ is used in the query) the point at which the
sanitization function call should be placed is determined as follows:\footnote{In our implementation, this is done simultaneously as part of the QFS algorithm, but this separate explanation is more understandable.}
\begin{enumerate}
\item Starting at node $n$ follow edges forward through the graph until a concatenation node is encountered.
\item Find the code location corresponding to this concatenation.
\item Insert a call to the sanitization function just before the concatenation is performed.
\end{enumerate}

\subsection{Implementation}

\begin{figure}
\includegraphics[scale=0.70]{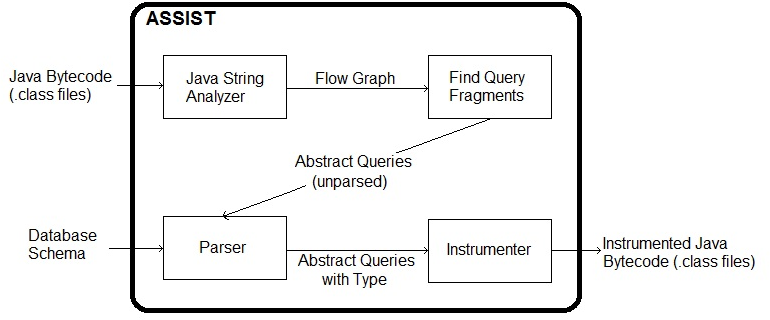}
\caption{Architecture of ASSIST}
\label{fig:ASSIST}
\end{figure}

Figure \ref{fig:ASSIST} shows the architecture of ASSIST. 
The main ideas of our approach are applicable to web applications written in a variety of languages. The ASSIST prototype is targeted to protect Java applications such as JSPs or Java Servlets; it takes Java bytecode as input and produces instrumented Java bytecode. ASSIST leverages the Java String Analyzer~\cite{strings2003} to construct the flow graph and uses Soot~\cite{soot} for further analysis and transformation of the bytecode.

The current prototype uses two sanitizing functions, which we provided.
The string sanitizing function checks for unescaped characters that could compromise the query, such as the quote ('), backslash (\textbackslash), etc. If any are found, the unescaped characters are sanitized by escaping them with backslash. It also checks for unescaped backslash at the end of the string. Note that the sanitizing function is designed to be able to work on top of existing ones without escaping the data twice. This is done by always looking ahead one character and checking whether the corresponding two character sequence is already escaped. The numeric sanitizing function attempts to parse the value as a number. If successful the sanitizing function returns the original value, or else if it fails the sanitization function returns the value \textit{null}, which will result in the value \textit{false} when the variable is being used to compare to any value in the query.

Other sanitizing functions could easily be substituted and, with small modifications to the prototype, special purpose sanitizers for domains other than string and numeric could be provided.

\section{Example Walk-through}

We now illustrate each step of our technique by showing a walk-through of running an example program through each phase of ASSIST. We show how our technique works with the code fragment in figure \ref{fig:code} and abstract the rest of the program away.  
There are three possible paths of execution through the program: the query will search for author, for price or an invalid query will be built.
ASSIST takes the compiled bytecode as input and invokes the Java String Analyzer~\cite{strings2003} to produce the flow graph in figure \ref{fig:flow_graph}.

ASSIST then performs the Find Query Fragments algorithm to find the possible abstract queries in the program. First it looks through each statement in the code for query execution points. They are the lines of code with an invocation to \textit{java.sql} library functions such as \textit{Statement.executeQuery(String)}. In our example line 11 contains a query execution point.
ASSIST locates its corresponding node from the flow graph, in this case, the node containing \textit{query}, and executes the Find Query Fragments algorithm on that node.
Since \textit{query} is on an Assignment node, the QFS of that node is the union of the QFS's from all its predecessors, i.e., 
the union of the QFS of the nodes representing statements \textit{query = r7}, \textit{query = r5.append(r6)}, and \textit{query = r10.append(r11)}.
First, consider
the node with \textit{query = r10.append(r11)}.
Since this is also an assignment node, its QFS is the QFS
of the concatenation node that is its predecessor.
The QFS of the concatenation node is the QFS from the left hand side concatenated with the QFS from the right hand side.
The QFS of the right hand side is the QFS of the node with \textit{r11 = getParam(“price”)};
the QFS of that node is the Initialization node containing the value \textit{any\_string}.
We use the target of the assignment, \textit{r11}, as the
place-holder in the abstract query.
Likewise, the left side of the concatenation node will lead to another concatenation node and the QFS of that node is \textit{SELECT * FROM BOOKS WHERE} concatenated with \textit{price $<$}.
So the QFS of the node with \textit{query = r10.append(r11)} is \textit{SELECT * FROM BOOKS WHERE price $<$ r11}.
In a similar manner, the other two paths produce \textit{SELECT * FROM BOOKS WHERE} and \textit{SELECT * FROM BOOKS WHERE author = 'r4'}.
Therefore the resulting abstract queries, or the QFS of the node containing \textit{query}, are the strings shown in figure \ref{fig:approximation_set}.

\begin{figure}
    \begin{tabular}{ | p{15cm} | }
    \hline
Line 11:\\
\sout{SELECT * FROM BOOKS WHERE}\\
SELECT * FROM BOOKS WHERE author = 'r4'\\
SELECT * FROM BOOKS WHERE price $<$ r11\\
    \hline

    \end{tabular}
  \caption{Abstract Queries of Example}
  \label{fig:approximation_set}
\end{figure}

After the abstract queries are computed, they are parsed.
The first query is an invalid query; we assume that it
was generated along an infeasible path. (Since it does not have
any host variables, it is not vulnerable to injection attacks, 
even if this path were feasible.)
By analyzing structures of the other queries and the database schema,
ASSIST determines that \textit{r4} is intended to be treated as a string in the query and \textit{r11} is intended to be treated as a number.
ASSIST then goes through the code and locates the lines of code right before the concatenation statements that adds \textit{r4} and \textit{r11} into their respective queries. ASSIST then instruments the code by inserting a call to the appropriate sanitizing function for each variable. The string sanitizing function is inserted for \textit{r4} and the numeric sanitizing function is inserted for \textit{r11}.

ASSIST outputs the instrumented bytecode. It is equivalent to the bytecode that would be generated by a modified version of the code in Figure~\ref{fig:code} with calls to the string sanitizer
on the value returned by {\tt getParameter("author")}
and a call to the numeric sanitizer 
on the value returned by {\tt getParameter("price")}.

\section{Limitations}
\label{sec:limitations}

There are several sources of imprecision which may lead to false positives and false negatives.
We did not encounter any of these cases in our experiments.

Our algorithm does not take into account infeasible paths in the application, leading to possible overapproximation of the set of abstract queries. This is however reduced by the parsing component where invalid SQL queries, generated along some infeasible paths, are eliminated.

Another source of imprecision is caused by cycles in the string flow graph,
which could arise from
programs that concatenate fragments onto queries in loops.
Such programs
have a potentially infinite number of abstract queries.
Our analysis only considers queries constructed through single iterations of loops in the flow graph, thus it underapproximates the set abstract queries.
Java String Analyzer~\cite{strings2003} provides a safe approximation of the set of queries, but abstracts away some of the information needed for our analysis. 
One direction for future research is modification of the FindQueryFragments algorithm to compute a safe approximation of the set of pairs of variables and the database attributes to which they correspond.

A third limitation is that it is possible to write a program which uses the same host variable as a string and a number in abstract queries.
For example, suppose that instead of providing separate parameter names, ``author'' and ``price'', the program in Figure~\ref{fig:code}
had a single parameter ``x'' and a single {\tt getParameter("x")} statement, before discriminating between the different possible actions.
Our analysis would show that the target variable of the {\tt getParameter} could be used as either a string or a number in the query.
In this case, our technique would not be able to determine its type to automatically apply the correct sanitization function; we can either sanitize the host variable as a number so it is safe for both fields (thereby limiting the application's functionality), or report this as an error so developers can check manually.

Currently, ASSIST only treats queries that are executed through {\it Statement} objects in Java. {\it PreparedStatement} objects, in which user inputs cannot modify the structure of the query, are also available in Java. {\it PreparedStatement} objects can provide protection against SQL injection but only if they are used correctly. A similar analysis to detect inappropriate usage of {\it PreparedStatement} objects can be a direction for future work. Also, other existing vulnerability detection techniques could potentially be used to determine that certain execution points are not vulnerable and to omit them from the analysis.

\section{Evaluation}

To evaluate ASSIST, we performed experiments to check whether the losses of precision discussed above lead to false positives (legitimate inputs that are identified as injection attacks) or false negatives (attacks that succeed) and to measure performance.
We expect the runtime overhead incurred by ASSIST to be low since only the instrumented sanitizing functions contribute to the runtime overhead, which is linear in the size of the user inputs and they are generally only several characters long.
We have conducted these experiments with the SQL Injection Application Testbed \cite{amnesia_testbed}, which was created to evaluate AMNESIA~\cite{amnesia}
and which has also been used for evaluating several other techniques~\cite{wasp,candid,essence,taint_inference}. It consists of a large number of test cases for a series of applications available at \textit{http://gotocode.com}. It contains two types of test cases: the ATTACK set which contains attempted SQL injection attacks, and the LEGIT set which contains legitimate inputs that look like SQL injection attacks. Table \ref{tbl:amnesia_set} summarizes this benchmark. The first column contains the names of the applications. The second column contains the number of lines of code (LOC) from each application. The remaining columns show the numbers of the different types of test cases from each set. All the applications are Java Server Pages.

The execution of the static analysis phase of ASSIST took about one minute per class file.
There are two types of test cases, an attack set and a legitimate set. The attack set is run to check whether ASSIST successfully prevents
attacks.
The legitimate set is run to determine whether any false positives reported by the tool; if the sanitizers modify inputs from the legitimate set, it would be a false positive. To help track what the sanitizers were doing, we modified ASSIST's sanitizing functions to output to a log every time they modified a user input.

We ran each test on the original application and on the instrumented code, produced by ASSIST and compared the database logs.
A difference between the queries executed by the original and by the instrumented code indicates an attack that has been stopped by ASSIST.
Logs of changes made by the sanitizers were examined to verify this.

As noted by Halfond, many of the inputs in the attack set are unsuccessful attacks~\cite{amnesia}.
Examination of the logs showed that all actual attacks were prevented by
ASSIST;
thus there were no false negatives.
None of the unsuccessful attacks and none of the legitimate inputs were modified by the
sanitizers, thus there were no false positives.

\begin{table}\scriptsize
  \begin{center}
    \begin{tabular}{ | c | c | c | c | c | c | c | }
      \hline
       & LOC & Cartesian & perParam & Random & Legit & Total \\
       &  & (ATTACK set) & (ATTACK set) & (ATTACK set) & (LEGIT set) & \\ \hline
      bookstore & 16,959 & 3063 & 410 & 2001 & 608 & 6082 \\ \hline
      classifieds & 10,949 & 3211 & 378 & 2001 & 576 & 6166 \\ \hline
      empldir & 5,658 & 3947 & 440 & 2001 & 660 & 7048 \\ \hline
      events & 7,242 & 3002 & 603 & 2001 & 900 & 6506 \\ \hline
      portal & 16,453 & 2968 & 717 & 2001 & 1080 & 6766 \\ \hline

      \end{tabular}
  \caption{Description of the SQL Injection Application Testbed}
  \label{tbl:amnesia_set}
  \end{center}
\end{table}

To determine the run time overhead of ASSIST we conduct timing experiments on the largest application in the testbed, the bookstore application.  We compared the runtime of the original bookstore application to that of the version that was instrumented by ASSIST. We ran the legitimate test set on them and measured the difference in execution time as overhead. We only used the legitimate test set for our timing experiments because the attack set would cause different paths of execution between the two versions, where attacks would be successful in the original application but be prevented in the instrumented application, leading to incorrect timing comparisons. To prevent network delay the applications are installed at localhost. To ensure accuracy we ran our timing experiments ten times and recorded the average run time. We found that the runtime overhead on the bookstore application is no more than 2\%.

\section{Related Work}

Many techniques have been developed to protect against SQL injection and other types of web applications injection attacks. The technique that is most similar to our technique of automatic query sanitization is PHP's magic quotes \cite{magic_quotes}. Like our technique it was intended as an automated sanitization measure against SQL injection. However unlike our technique it does so by blindly escaping all quote characters from the user input without the use of any static analysis. Simply escaping quotes however turns out to be a poor measure against SQL injection, and doing so on every user input causes issues when constructing statements of other languages like HTML and Javascript as well. In the end it caused more problems than it intended to solve and is currently being removed from the language altogether. Our technique avoids these limitations with the use of its underlying static analysis technique and the instrumentation of more sophisticated sanitization functions than just escaping quotes.

AMNESIA~\cite{amnesia} is another technique that is similar to ASSIST, in that both use a combination of string analysis and code instrumentation.
AMNESIA uses the Java String Analyzer to construct an automaton representing the structures (command forms) of expected queries,
then inserts run-time monitors at query execution points. The run-time monitors check whether queries that are about to be sent to the DBMS match the automaton to detect occurrences of attacks.
While AMNESIA checks that queries constructed have expected structures,
ASSIST
adds sanitizing functions in order to prevent unexpected queries from being
generated.
In terms of implementation,
AMNESIA uses the automaton output of the Java String Analyzer, while ASSIST
analyzes JSA's internal flow graph. 

Static techniques \cite{su_wass_static, java_static, pixy, huang_static, saner, taj} generally employ the use of various forms of static code analysis to locate sources of injection vulnerabilities in code. The difference between these techniques and ASSIST is that while other techniques employ static analysis to detect vulnerable code or occurrences of attacks, ASSIST uses static analysis to find host variables and automatically sanitizes them by instrumenting them with calls to sanitization functions. 

Machine learning techniques \cite{waves, learning} involve finding SQL injection vulnerabilities through the use of training sets. Martin, Livshits, and Lam developed PQL \cite{pql}, a program query language that developers can use to find answers about injection flaws in their applications and suggested that static and dynamic techniques can be developed to solve these queries.

Dynamic tainting techniques \cite{wasp, php_taint, csse, java_taint, efficient_java_taint, c_taint, taint_inference} are runtime techniques which generally involve the idea of marking every string within a program with taint variables and propagating them across execution. Attacks are detected when a tainted string is used as a sensitive value. Bandhakavi, Bisht, Madhusudan, Venkatakrishnan developed CANDID \cite{candid}, where candidate clones of every string are created and propagated across execution. Eventually two versions of a SQL query are executed: the original query with user inputs and a candidate version with some benign value in place of user inputs, which are compared to determine if an attack occurred. Buehrer, Weide, and Sivilotti developed a technique involved with comparing parse trees \cite{parse_tree} to prevent SQL injection attacks. Su and Wassermann also developed an approach involving parse trees \cite{essence} that marks sections of the input that come from users and checks whether they occur in contexts that meet a language-based security policy.

Boyd and Keromytis developed a technique called SQLrand \cite{sqlrand} to prevent SQL injection attacks based on instruction set randomization. SQL keywords are randomized at the database level so attacks from user input become syntactically incorrect SQL statements. A proxy is set up between the web server and the database to perform randomization of these keywords using a key.

\section{Conclusion and Future Work}

In this paper, we have presented the ASSIST technique for automatic query sanitization. ASSIST uses a combination of static analysis and program transformation to automatically locate and perform sanitization on host variables that are used to construct SQL queries. We have implemented our technique in Java with a tool named ASSIST for protecting Java bytecode (derived from JSPs or Servlets).
An empirical evaluation  demonstrated that ASSIST is effective against an SQL injection attack test suite and that ASSIST operates with a very low runtime overhead. 
Directions for future work include more extensive evaluation, including direct comparison with other techniques; modification of the algorithms to reduce potential imprecision; and extending the technique to check other properties of user inputs, such as conformance with the domain of database table attributes to which they are assigned or compared. Although the experiments indicate that the algorithm is precise enough for our purposes, improvements in the precision to eliminate the possibility of false negatives might be worthwhile.

\section{Acknowledgments}

This research was partially supported by the US Department of Education GAANN grant P200A090157, National Science Foundation grant CCF 0541087, and the Center for Advanced Technology in Telecommunications sponsored by NYSTAR.


\bibliographystyle{plain} 
\bibliography{my_bib}

\end{document}